# Dissolution of sulfur dioxide and nitrogen monoxide in water

Fernando Hevia[a,*], Barbara Liborio[b], Karine Ballerat-Busserolles [b], Yohann Coulier[b], Jean-Yves Coxam[b]

[a] G.E.T.E.F., Departamento de Física Aplicada, Facultad de Ciencias, Universidad de Valladolid. Paseo de Belén, 7, 47011 Valladolid, Spain.

[b] Université Clermont Auvergne, CNRS, SIGMA Clermont, Institut de Chimie de Clermont-Ferrand, F-63000, Clermont-Ferrand, France

[*] e-mail: fernando.hevia@termo.uva.es

**Abstract**

Sulfur dioxide ($SO_2$) and nitrogen monoxide (NO) are some of the gas impurities present in the carbon dioxide ($CO_2$) separated from fume using post combustion capture process. Even in a small amount, these impurities may have an impact on storage process development. The dissolution of such gases in aqueous phase is part of the studies carried out to develop processes and analyze the risks in case of geological storage. The enthalpies of solution of $SO_2$ and NO in water were here investigated by calorimetry, using a flow calorimetric technique. The enthalpies of solution were determined at 323.15 K and 373.15 K for sulfur dioxide, and at 323.15 K for nitrogen monoxide. The experimental enthalpy data were used together with available solubility data to test thermodynamic models representative of vapor-liquid equilibrium.

**Keywords**: enthalpy of solution; sulfur dioxide; nitrogen monoxide; water; solubility; flow calorimetry.

# 1. Introduction

Carbon dioxide ($CO_2$) removal from post-combustion industrial effluents could contribute substantially to the reduction of anthropogenic emission of carbon [1]. Indeed, $CO_2$ capture processes are developed to mitigate industrial emissions and tackle climate change issues. The recovered carbon dioxide will be then stored in a secure site, such as deep saline aquifers. However, depending on the effluent origin [2] and the selected capture process, the injected $CO_2$ stream may contain small amounts of associated gaseous components such as $O_2$, $N_2$, $SO_x$, $NO_x$. These impurities will impact the thermo-physical properties (density, viscosity) and phase diagram behavior of the gas mixture [3] and then, co-injection of such gas impurities with $CO_2$ will have to be taken into consideration for the development of geological storage processes [4].

Before looking at the effects of the co-injection of multiple gas impurities with $CO_2$ in geological fluids, we first focus on the dissolution of $SO_2$ in water and NO in water. The aim of the paper is to select a model capable of predicting the gas solubility and the enthalpy of solution as functions of temperature and pressure of the {$SO_2$-$H_2O$} and {NO-$H_2O$} systems. This model will represent the vapor-liquid and chemical equilibria taking into account the non-ideality in liquid and gas phases. Because these are the most available experimental data, the thermodynamic models are generally adjusted with solubility data. Then the enthalpies are essentially derived from solubility data using Van't Hoff relations. Direct measurements of solution enthalpies will make possible to test the robustness and consistency of the models.

Literature data for the enthalpy of solution of $SO_2$ in water are scarce. The enthalpy of solution was experimentally determined at 298.15 K as a function of loading charge (ratio of the overall amount of substance of $SO_2$ to that of $H_2O$) by Stiles and Felsing [5]. Johnstone and Leppla [6] determined ionization constants of sulfurous acid and Henry's law coefficients at temperatures from 273.15 K to 323.15 K. The authors report enthalpies of ionization and enthalpies of solution, derived from the temperature dependence of the ionization constant and Henry's law constant, respectively. No enthalpy data were found for the dissolution of NO in water.

In this work, we determine experimentally the enthalpies of solution of $SO_2$ and NO in water at different temperatures and pressures. Details of the experimental conditions are reported in Table 1. The measurements were carried out using a flow calorimetry technique previously developed to study hydrogen sulfide ($H_2S$) dissolution in aqueous solutions [7]. We also represent the vapor-liquid equilibrium of the {$SO_2$-$H_2O$} system using the thermodynamic model developed by Rumpf and Maurer [8]. The model was assessed using available experimental solubility data [6,8-21]. A calculation of the enthalpies of solution, based on a derivation of the Gibbs energy of solution, has been added to this model to test its ability to represent both solubility and enthalpy data. On the other hand, the {NO-$H_2O$} system was considered to be ideal because of a very low solubility of NO, and vapor-liquid equilibrium was represented by Henry's law. Experimental vapor-liquid equilibrium data of NO in water [22,23] were used to calculate Henry's law constant, from which the enthalpy of solution can be derived.

Table 1. Temperature ($T$) and average pressure ($<p>$) of the enthalpy of solution measurements of gas in water.

| Gas | $T$ / K | $<p>$ / MPa |
|---|---|---|
| $SO_2$ | 323.15, 373.15 | 0.31 |
| NO | 323.15 | 2.27, 2.56, 2.80 |

## 2. Experimental

### 2.1. Material

Details about the source and purity of the investigated gases can be found in Table 2. They were used without further purification. Water (CAS: 7732-18-5) was triple distilled and degassed under vacuum.

Table 2. Sample description.

| Chemical name | CAS | Source | Purification method | Mole fraction purity[a] |
|---|---|---|---|---|
| Sulfur dioxide ($SO_2$) | 7446-09-5 | Linde (France) | None | ≥ 0.998 |
| Nitrogen monoxide (NO) | 10102-43-9 | Linde (France) | None | ≥ 0.998 |

[a] Provided by the supplier.

### 2.2. Apparatus and procedure

The flow calorimetric technique (Fig. 1) is similar to the one used by Koschel *et al*. [7] for the measurement of the enthalpies of solution of hydrogen sulfide ($H_2S$) in water. The principle consists of flowing gas and water from two high pressure syringe pumps into a mixing cell, located inside a Calvet-type calorimeter (Setaram C80). The heat power effect during dissolution is detected by a thermopile surrounding the mixing cell. The temperature of the calorimetric block is kept constant by a Setaram G11 Universal Controller with a standard uncertainty of 0.03 K. The pressure is read along the flow line by pressure gauges located at the input and output of the mixing cell (Fig. 1), and its standard uncertainty is 0.02 MPa. The difference between the pressures measured by these gauges is lower than the standard uncertainty on pressure.

The detected heat power ($\dot{q}$) is proportional to the electromotive force of the thermopile (S), $\dot{q} = K \cdot S$. The temperature-dependent calibration constant (K) is determined by chemical calibration using the reference system {ethanol-water} [24,25]. The enthalpy of solution per mole of substance $i$ ($i$ = w for water, and $i$ = $SO_2$ or NO for the gas) is calculated from the eq. 1, by dividing the heat power by the gas or water molar flow rate, $\dot{n}_i$, which are obtained from the volume flow rates ($\dot{V}_i$) and fluid molar volumes ($V_i$).

$$\Delta_{sol} H_i = \frac{\dot{q}}{\dot{n}_i} = \frac{\dot{q} \cdot V_i}{\dot{V}_i} \qquad (i = \text{w, } SO_2 \text{ or NO}) \qquad (1)$$

The molar volumes of $SO_2$ and water were taken from the equations of state recommended by NIST [26,27]. The molar volume of NO was calculated using the translated-consistent-Peng-Robinson equation of state as described by Le Guennec *et al.* [28].

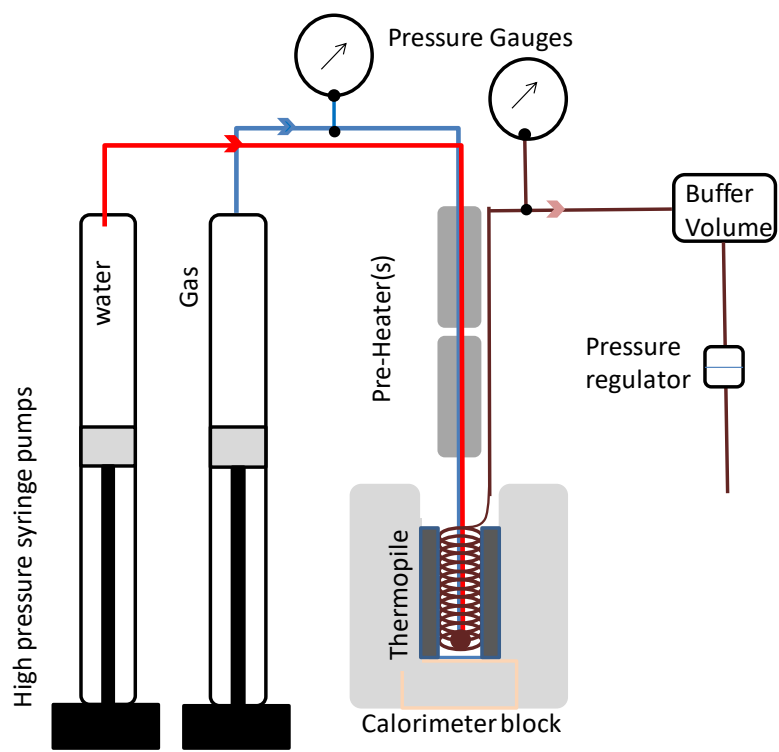

Figure 1. Schematic representation of the flow calorimetric technique.

The enthalpies of solution of $SO_2$ and NO in water are determined as functions of gas loading charge $\alpha = \dot{n}_{gas}/\dot{n}_w$, which is the ratio of gas (=$SO_2$ or NO) molar flow rate ($\dot{n}_{gas}$) and water molar flow rate ($\dot{n}_w$). The loading charge is related to the overall molality of the gas (i.e., before dissociation; see below), $\bar{m}_{gas} = \alpha/M_w$, where $M_w$ is the molar mass of water.

The enthalpy of solution is expressed either per mole of water ($\Delta_{sol}H_w$) or per mole of gas ($\Delta_{sol}H_{gas}$). The representation of the enthalpy of solution per mole of water as a function of the gas loading charge is used for graphical estimation of the gas solubility. Indeed, $\Delta_{sol}H_w$ increases up to reach a plateau when the solution is gas-saturated. The limit of gas solubility corresponds to the intersection between the part of the curve where the enthalpy increases and the plateau. For a correct determination, the measurements must be carried out below and above the limit of gas solubility. For small gas solubility, the measurements below saturation are restricted by the difficulties to control gas dissolution with small gas volume flow rates. In addition, small gas absorptions are associated to small heat effects, and then high relative uncertainties on the calorimetric signal. In contrast, for systems with high gas solubility limits, the experimental difficulties are encountered for the measurements in the saturated domain. For a gas loading charge above the limit of solubility, only a part of the gas is absorbed and gas bubbles appear in the mixing cell. Furthermore, for these measurements in the saturated domain high flow rates are required, and thus the residence time inside the mixing cell may become too short for a complete heat exchanged via the thermopile.

## 3. {$SO_2$-$H_2O$} system

### 3.1. Thermodynamic equations for chemical and physical equilibria

The non-ideality of the liquid phase is described by means of activity coefficients ($\gamma_i$ for species $i$). For liquid water (solvent), $\gamma_w$ is mole-fraction based and normalized to one for pure water. The $\gamma_i$ of the rest of the species in the liquid phase (solutes) are based on molalities ($m_i$), and are normalized to one at infinite dilution in the solvent.

The representation of the dissolution of $SO_2$ in water implies accounting for the presence of chemical reactions in the liquid phase. The chemical reactions of $SO_2$ in water [8,17,29] are given in eq. I for first $SO_2$ dissociation, eq. II for second $SO_2$ dissociation; water dissociation is represented by eq. W.

$$SO_2 + H_2O \rightleftharpoons HSO_3^- + H^+ \tag{I}$$

$$HSO_3^- \rightleftharpoons SO_3^{2-} + H^+ \tag{II}$$

$$H_2O \rightleftharpoons H^+ + OH^- \tag{W}$$

The equations for chemical equilibria are given by the law of mass action (eq. 2) where $a_i$ and $v_i$ are the activity and the algebraic stoichiometric coefficient of species $i$ in reaction $N$, respectively. The equilibrium constants $K_N$, defined according to the standard states described above, are represented as functions of temperature following eq. 3; values of coefficients $A_N$, $B_N$, $C_N$ and $D_N$ are summarized in Table 3.

$$K_N = \prod_i a_i^{v_i} \qquad \text{with } N = \text{I, II or W} \qquad (2)$$

$$\ln K_N = \frac{A_N}{(T/\text{K})} + B_N \cdot \ln(T/\text{K}) + C_N \cdot (T/\text{K}) + D_N \qquad (3)$$

Table 3. Parameters used in eq. 3 to represent the temperature dependence of equilibrium constants $K_N$.

| Reaction $N$ | $A_N$ | $B_N$ | $C_N$ | $D_N$ | Reference |
|---|---|---|---|---|---|
| I | 26404.29 | 160.3981 | –0.2752224 | –924.6255 | [30] |
| II | –5421.930 | 4.689868 | –0.04987690 | 43.13158 | [30] |
| W | –13445.9 | –22.4773 | 0 | 140.932 | [29] |

The mass balance equations for the overall amounts $\bar{n}_i$ of sulfur dioxide and water are represented by eqs. 4-5, where $n_i$ is the amount of component $i$ in the liquid phase at equilibrium.

$$\bar{n}_{SO_2} = n_{SO_2} + n_{HSO_3^-} + n_{SO_3^{2-}} \qquad (4)$$

$$\bar{n}_w = n_w + n_{HSO_3^-} + n_{SO_3^{2-}} + n_{OH^-} \qquad (5)$$

The charge balance is represented by eq. 6.

$$n_{H^+} = n_{HSO_3^-} + 2n_{SO_3^{2-}} + n_{OH^-} \qquad (6)$$

The vapor-liquid equilibria are defined by eqs. 7-8. All ionic species have been considered as non-volatile.

$$H_2O^{(l)} \rightleftharpoons H_2O^{(v)} \qquad (7)$$

$$SO_2^{(l)} \rightleftharpoons SO_2^{(v)} \qquad (8)$$

The conservation equation in the vapor phase is expressed by eq. 9, where $y_i$ is the mole fraction of component $i$ in the vapor.

$$y_{SO_2} + y_w = 1 \qquad (9)$$

The equations representative of the vapor-liquid equilibrium at temperature $T$ and pressure $p$ are stated using extended Raoult's law (eq. 10) for water and extended Henry's law for $SO_2$ (eq. 11).

$$\phi_w y_w p = p_w^s \phi_w^s \exp\left[\frac{V_w(p - p_w^s)}{RT}\right] a_w \qquad (10)$$

$$\phi_{SO_2} y_{SO_2} p = H_{SO_2,w} \exp\left[\frac{\bar{V}_{SO_2}^\infty (p - p_w^s)}{RT}\right] m_{SO_2} \gamma_{SO_2} \qquad (11)$$

The saturation pressure ($p_w^s$) and molar volume ($V_w$) of water were taken from recommended equations given by Saul and Wagner [31]. The partial molar volume of $SO_2$ at infinite dilution in water ($\bar{V}_{SO_2}^\infty$) was calculated as a function of temperature (eq. 12) using a simple second-order polynomial equation adjusted on values obtained from Brelvi and O'Connell [32] and given in reference [8].

$$\bar{V}_{SO_2}^\infty / cm^3 \cdot mol^{-1} = 84.2113 - 0.334941 \cdot (T/K) + 6.35086 \cdot 10^{-4} (T/K)^2 \qquad (12)$$

The molality-based Henry's law constant of $SO_2$ in water at the saturation pressure of water ($H_{SO_2,w}$, eq. 13) is calculated using the correlation developed by Rumpf and Maurer [8].

$$\ln\left[H_{SO_2,w} / (MPa \cdot kg \cdot mol^{-1})\right] = -154.827 + \frac{321.17}{(T/K)} \qquad (13)$$
$$-0.0634 \cdot (T/K) + 29.872 \cdot \ln(T/K)$$

The vapor phase is represented by a virial equation of state (eq. 14), truncated after the second virial coefficient ($B$).

$$\frac{pV}{RT} = 1 + \frac{Bp}{RT} \qquad (14)$$

The second virial coefficient of the mixture ($B$), is calculated (eq. 15) from the virial coefficients of the pure components, $B_{ii}$ ($i$ = w, $SO_2$), and the symmetric cross coefficients $B_{ij}$ ($i,j$ = w, $SO_2$; and $i \neq j$). The second virial coefficient of pure water was calculated (eq. 16) from the correlation of Bieling *et al.* [33].

$$B = \sum_i \sum_j y_i y_j B_{ij} \qquad (15)$$

$$B_{w,w} / cm^3 \cdot mol^{-1} = -53.53 - 39.29 \cdot \left[\frac{647.3}{(T/K)}\right]^{4.3} \qquad (16)$$

Mixed second virial coefficients $B_{ij}$ are calculated with the method proposed by Hayden and O'Connell [34]. Pseudocritical temperatures and pressures ($T_{c,i}$, $p_{c,i}$), molecular dipole moments ($\mu_i$), and mean radii of gyration ($R_{D,i}$) of the pure components as well as association parameters ($\eta_{ij}$) were taken from reference [34] (Table 4).

Table 4. Parameters for the Hayden and O'Connell equation for estimating pure and mixed second virial coefficients.

(a) Pure component parameters: critical temperature ($T_{c,i}$), critical pressure ($p_{c,i}$), dipole moment ($\mu_i$) and radius of gyration ($R_{D,i}$).

|   | $T_{c,i}$ / K | $p_{c,i}$ / MPa | $\mu_i$ / D | $R_{D,i}$ / $10^{-10}$ m |
|---|---|---|---|---|
| $H_2O$ | 647.3 | 22.13 | 1.83 | 0.615 |
| $SO_2$ | 430.7 | 7.78 | 1.51 | 1.674 |

(b) Value of the parameter $\eta_{ij}$ for association between molecules $i$ and $j$.

|   | $j = H_2O$ | $j = SO_2$ |
|---|---|---|
| $i = H_2O$ | 1.7 | 0 |
| $i = SO_2$ | 0 | 0 |

The fugacity coefficients of the components are calculated using eq. 17.

$$\ln \phi_i = \frac{p}{RT}\left(2\sum_j y_j B_{ij} - B\right) \qquad (17)$$

The activity coefficients ($\gamma_i$) of the solutes in the liquid phase are calculated (eq. 18) using the Pitzer model as modified by Edwards *et al.* [29], where $\alpha = 2.0$ kg$^{1/2}$·mol$^{-1/2}$, $b = 1.2$ kg$^{1/2}$·mol$^{-1/2}$, $z_i$ is the charge number of species $i$, $A_\phi$ is the Debye-Hückel limiting slope for the osmotic coefficient, and $I$ is the ionic strength (eq. 19).

$$\begin{aligned}
\ln \gamma_i = &-A_\phi z_i^2 \left[\frac{\sqrt{I}}{1+b\sqrt{I}} + \frac{2}{b}\ln\left(1+b\sqrt{I}\right)\right] \\
&+ 2\sum_{j\neq w} m_j \left\{\beta_{ij}^{(0)} + \beta_{ij}^{(1)} \frac{2}{\alpha^2 I}\left[1-\left(1+\alpha\sqrt{I}\right)\exp\left(-\alpha\sqrt{I}\right)\right]\right\} \\
&- z_i^2 \sum_{j\neq w}\sum_{k\neq w} m_j m_k \beta_{jk}^{(1)} \frac{1}{\alpha^2 I^2}\left[1-\left(1+\alpha\sqrt{I}+\frac{\alpha^2 I}{2}\right)\exp\left(\alpha\sqrt{I}\right)\right] \\
&+ 3\sum_{j\neq w}\sum_{k\neq w} m_j m_k \tau_{ijk}
\end{aligned} \qquad (18)$$

$$I = \frac{1}{2}\sum_{i\neq w} m_i z_i^2 \qquad (19)$$

$A_\phi/\mathrm{kg}^{1/2}\cdot\mathrm{mol}^{-1/2}$ is calculated as a function of the temperature using the equation given by Chen *et al.* [35].

$\beta_{ij}^{(0)}$, $\beta_{ij}^{(1)}$ and $\tau_{ijk}$ are binary zeroth-order, binary first-order and ternary interaction parameters among the solute species respectively. The activity of water is given by eq. 20.

$$\ln a_\mathrm{w} = M_\mathrm{w} \left\{ 2A_\phi \frac{I^{3/2}}{1+b\sqrt{I}} - \sum_{i\neq \mathrm{w}}\sum_{j\neq \mathrm{w}} m_i m_j \left[ \beta_{ij}^{(0)} + \beta_{ij}^{(1)} \exp\left(-\alpha\sqrt{I}\right) \right] \right.$$
$$\left. -2\sum_{i\neq \mathrm{w}}\sum_{j\neq \mathrm{w}}\sum_{k\neq \mathrm{w}} m_i m_j m_k \tau_{ijk} - \sum_{i\neq \mathrm{w}} m_i \right\} \quad (20)$$

Following Rumpf and Maurer's approach [8], parameters for interactions between molecular and ionic species and those between ionic species themselves were neglected, because the majority of dissolved $SO_2$ is molecular. For molecular $SO_2$-$SO_2$ interactions, only the binary zeroth order parameter $\beta_{SO_2,SO_2}^{(0)}$ (eq. 21) was retained.

$$\beta_{SO_2,SO_2}^{(0)} / \left(\mathrm{kg}\cdot\mathrm{mol}^{-1}\right) = 0.0934 - \frac{137.92}{(T/\mathrm{K})} + \frac{3.127\cdot 10^4}{(T/\mathrm{K})^2} \quad (21)$$

In order to calculate the equilibrium concentrations given two of the variables of the set $(T, p, \bar{m}_{SO_2})$, where $\bar{m}_{SO_2}$ is the overall molality of $SO_2$, the 9 following equations were solved simultaneously using the iterative Newton-Raphson technique: mass balance in the solution (eqs. 4-5) and in the gas phase (eq. 9), charge balance (eq. 6), chemical equilibria (eq. 2) for all the reactions involved in $SO_2$ dissolution (I, II and W) and vapor-liquid equilibria (eqs. 10-11).

### 3.2. Thermodynamic calculation of the enthalpy of the process

All the quantities of the processes in this paragraph will refer to one mole of dissolved $SO_2$. The enthalpy of solution ($\Delta_\mathrm{sol} H_{SO_2}$) is the sum of the enthalpy of physical dissolution ($\Delta_\mathrm{r,8} H$) and the enthalpies $\Delta_{\mathrm{r},N} H$ of all the chemical reactions (eq. 22).

$$\Delta_\mathrm{sol} H_{SO_2} = \Delta_\mathrm{r,8} H + \sum_N \Delta_{\mathrm{r},N} H \quad (22)$$

To calculate the enthalpy of physical dissolution $\Delta_\mathrm{r,8} H$, we consider the Gibbs energy change of the vapor-liquid equilibrium of $SO_2$ (eq. 23) and use the Gibbs-Helmholtz relation (eq. 24).

$$\frac{\Delta_\mathrm{r,8} G}{RT} = \ln H_{SO_2,\mathrm{w}} + \ln \gamma_{SO_2} + \frac{\bar{V}_{SO_2}^\infty \left(p - p_\mathrm{w}^\mathrm{s}\right)}{RT} + \ln m_{SO_2} - \ln \phi_{SO_2} - \ln\left(y_{SO_2} p\right) \quad (23)$$

$$\Delta_{r,8}H = -RT^2 \left( \frac{\partial}{\partial T} \left\{ \frac{\Delta_{r,8}G}{RT} \right\} \right)_{p,\{m_i\}}$$
$$= -RT^2 \left( \frac{\partial}{\partial T} \left\{ \ln H_{SO_2,w} + \ln \gamma_{SO_2} + \frac{\overline{V}_{SO_2}^{\infty}(p - p_w^s)}{RT} - \ln \phi_{SO_2} \right\} \right)_{p,\{m_i\}} \quad (24)$$

The enthalpy of reaction $N$ ($N$ = I, II, W) per mole of $SO_2$ (eq. 25) is calculated from the standard enthalpy of reaction ($\Delta_{r,N}H^\circ$) (eq. 26), the excess partial molar enthalpy of every compound $i$ ($\overline{H}_i^E$) (eq. 27), and the extent of reaction ($\xi_N$) when one mole of gas is absorbed.

$$\Delta_{r,N}H = \frac{\xi_N}{\overline{n}_{SO_2}} \left( \Delta_{r,N}H^\circ + \sum_i \nu_{i,N} \overline{H}_i^E \right) \quad (25)$$

$$\Delta_{r,N}H^\circ = RT^2 \left( \frac{\partial \ln K_N}{\partial T} \right)_p \quad (26)$$

$$\overline{H}_i^E = -RT^2 \left( \frac{\partial \ln \gamma_i}{\partial T} \right)_{p,\{m_i\}} = -RT^2 \left( \frac{\partial \ln a_i}{\partial T} \right)_{p,\{m_i\}} \quad (27)$$

### 3.3. Results and discussion

*3.3.1. Model validation*

The thermodynamic model, taken into account the non-ideality (§3.1), was assessed using gas solubility data from literature [6,8-21]. The relative standard deviations, $\sigma_r(\overline{m}_{SO_2})$, between calculated and experimental solubility (molality) data have been determined for each reference and are given in Table 5. The model provides a satisfactory representation of most of the experimental solubility data with a relative standard deviation around 5%. A significant scattering is observed between some of the oldest references, published before 1939 [13,15,18], and the rest of the data. The high relative standard deviation observed with Shaw *et al.* [17] data concerns only 3 data points for which the authors reported possible experimental imprecision. The relative deviation, $\delta = (\overline{m}_{SO_2}^{calc} - \overline{m}_{SO_2}^{exp})/\overline{m}_{SO_2}^{exp}$, of the predicted solubility ($\overline{m}_{SO_2}^{calc}$) from the experimental data ($\overline{m}_{SO_2}^{exp}$) issued from the rest of the mentioned references, is plotted against experimental pressure in Figure 2. The graphs show that the relative deviations are symmetrically distributed around $\delta = 0$. The results of Figure 2a concern only the data from Rumpf and Maurer [8] and we have checked that our computations, determined with the set of parameters from this reference, are compatible with the calculations of the authors [8].

Table 5. Experimental solubility data of gaseous $SO_2$ in liquid water, with their temperature ($T$) and pressure ($p$) ranges, used to test the validity of the model [8].

| Reference | Data points | $T$ / K | $p$ / MPa | $100\sigma_r(\bar{m}_{SO_2})$ [a] |
|---|---|---|---|---|
| Beuschlein and Simenson [9] | 53 | 296.35 to 386.15 | 0.0653 to 1.925 [b] | 6.7 |
| Byerley et al. [10] | 2 | 298.15 to 323.15 | 0.101325 [c] | 5.0 |
| Douabul and Riley [11] | 6 | 278.97 to 303.25 | 0.101325 [c] | 7.8 [d] |
| Hudson [12] | 42 | 283.15 to 363.15 | 0.1008 to 0.1272 [c] | 3.9 |
| Johnstone and Leppla [6] | 16 | 298.15 to 323.15 | 0.000027 to 0.001370 [b] | 4.3 |
| Maass and Maass [13] | 29 | 283.15 to 300.15 | 0.0324 to 0.3408 [c] | 18.3 |
| Mondal [14] | 20 | 293 to 333 | 0.000447 to 0.000963 [b] | 4.7 [d] |
| Otuka [15] | 22 | 373 to 423 | 0.174 to 0.354 [c] | 29.7 [d] |
| Rabe and Harris [16] | 43 | 303.15 to 353.15 | 0.00523 to 0.101 [b] | 3.8 |
| Rumpf and Maurer [8] | 66 | 293.14 to 393.33 | 0.0356 to 2.509 [c] | 2.5 |
| Shaw et al. [17] | 3 | 297.75 to 312.25 | 0.0252 to 0.196 [c] | 43.6 |
| Sherwood [18] | 109 | 273.15 to 323.15 | 0.000033 to 0.0968 [b] | 30.4 |
| Siddiqi et al. [19] | 50 [e] | 290.15 to 294.65 | 0.0000011 to 0.000534 [b] | 5.4 |
| Smith and Parkhurst [20] | 8 | 278.15 to 333.15 | 0.02342 to 0.14769 [b] | 3.3 |
| Tokunaga [21] | 4 | 283.15 to 313.15 | 0.1025 to 0.1087 [c] | 4.6 |

[a] $\sigma_r(\bar{m}_{SO_2}) = \sqrt{\dfrac{1}{N_p - 1} \sum_i \left( \dfrac{\bar{m}_{SO_2,i}^{calc} - \bar{m}_{SO_2,i}^{exp}}{\bar{m}_{SO_2,i}^{exp}} \right)^2}$, relative standard deviations from the model; $N_p$, number of experimental data points; $\bar{m}_{SO_2,i}^{exp}$, experimental solubility; $\bar{m}_{SO_2,i}^{calc}$, calculated solubility. [b] Partial pressure of $SO_2$. [c] Total pressure. [d] Molality estimated from molarity using density of water. [e] Excluded two points reported with zero partial pressure of $SO_2$.

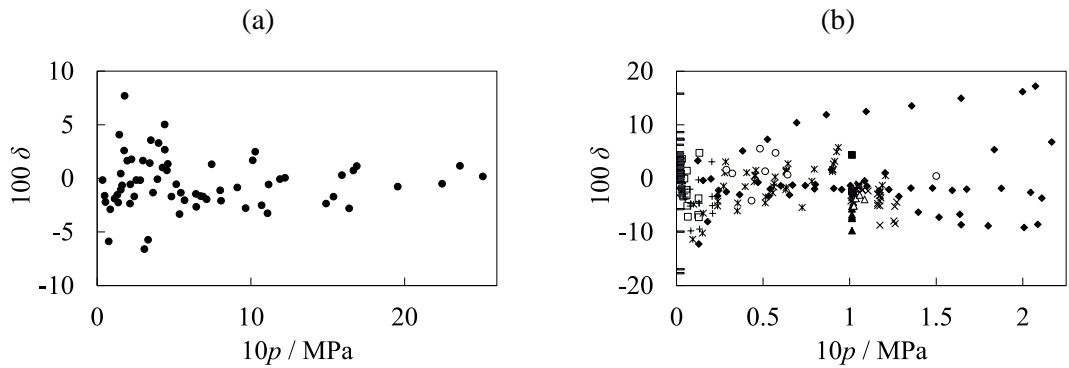

Figure 2. Relative deviation, $\delta$, of the predicted from the experimental solubility (molality) of gaseous sulfur dioxide in liquid water as a function of experimental pressure, $p$. (a) Data from Rumpf and Maurer [8]. (b) Data from other references. (♦): Beuschlein and Simenson [9]; (■): Byerley et al. [10]; (▲): Douabul and Riley [11]; (×): Hudson [12]; (□): Johnstone and Leppla [6]; (+): Mondal [14]; (*): Rabe and Harris [16]; (—): Siddiqi et al. [19]; (○): Smith and Parkhurst [20]; (△): Tokunaga [21].

The enthalpies of solution derived from the vapor-liquid equilibrium model were compared to the experimental data at 298.15 K by Stiles and Felsing [5]. As illustrated in Figure 3, the model adequately represents the experimental enthalpies.

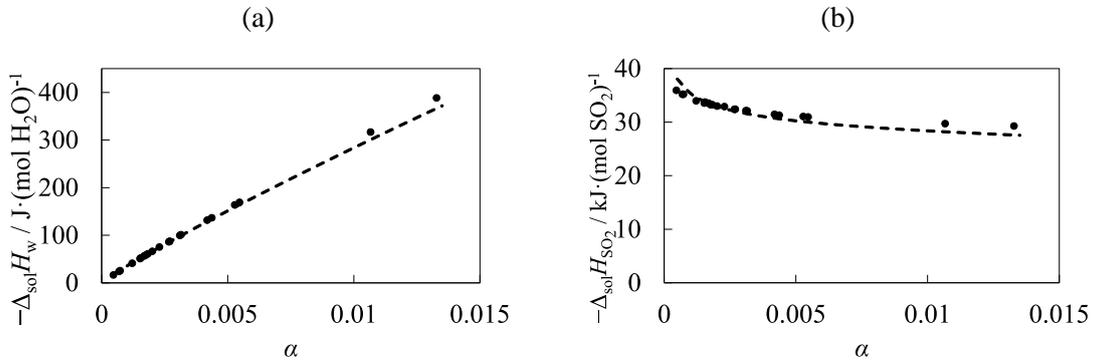

Figure 3. Enthalpy of solution of gaseous sulfur dioxide in liquid water per mole of water (a), $\Delta_{sol}H_w$, and per mole of sulfur dioxide (b), $\Delta_{sol}H_{SO_2}$, at temperature $T$ = 298.15 K. (●): experimental values (Stiles and Felsing [5]); (- - -): calculated values.

*3.3.2. Results*

The calorimetric measurements were carried out at temperatures $T$ = 323.15 K and $T$ = 373.15 K and pressure around an average value of 0.31 MPa. Experimental values are reported in Table 6. The experimental points with a loading charge above the theoretical gas solubility limit derived from the model are considered to correspond to a two-phase system. In the gas-saturated region, the loading charge and overall molality are then referred to the total composition of the two-phase system. The experimental and calculated enthalpies of solution are plotted as functions of gas loading charge in Figure 4.

At 323.15 K, $\Delta_{sol}H_w$ increases with gas loading charge ($\alpha$) up to the limit of solubility. Above the calculated solubility limit ($\alpha_{calc}$ = 0.035 at $p$ = 0.31 MPa), the enthalpy remains constant. However, the number of experimental points in the saturated domain (Figure 4a) is not sufficient to make possible an accurate experimental estimation of the solubility limit. The highest gas loading charge $\alpha$ was obtained with sulfur dioxide and water volume flow rates of 0.85 mL/min and 0.05 mL/min, respectively. These experimental flow rates correspond to the upper and lower limits for the gas and the liquid, respectively. Concerning the thermodynamic model, the calculated enthalpies of solution are in good agreement with the experimental data as shown in Figures 4a and 4b.

At 373.15 K, the gas solubility is low enough to allow measurements in the saturated domain (Figure 4c). The calculated solubility limit of $SO_2$ in water ($\alpha_{calc}$ = 0.008 at $p$ =0.31 MPa), is consistent with the value which can be determined from the experimental enthalpy of solution $\Delta_{sol}H_w$ (see Figure 4c). A correct representation of the enthalpy of solution (Figures 4c and 4d) is obtained from the model for the lowest gas loading but enthalpies are overestimated when approaching the gas solubility limit.

Table 6. Enthalpy of solution of gaseous sulfur dioxide in liquid water per mole of water, $\Delta_{sol}H_w$, and per mole of sulfur dioxide, $\Delta_{sol}H_{SO_2}$, at temperatures $T$ = 323.15 K and $T$ = 373.15 K and pressure $p$ (final pressure after mixing), as functions of the loading charge of the gas (moles of gas per 1 mole of water), $\alpha = M_w \bar{m}_{SO_2}$ ($M_w$, molar mass of water; $\bar{m}_{SO_2}$, overall molality of $SO_2$.). $u(X)$, standard uncertainty of the quantity $X$; $U(X)$ expanded uncertainty (95% confidence level) of the quantity $X$. [a]

| $p$ /MPa | $\alpha$ | $u(\alpha)$ | $\bar{m}_{SO_2}$ /mol·kg$^{-1}$ | $-\Delta_{sol}H_w$ /J·(mol H$_2$O)$^{-1}$ | $U(\Delta_{sol}H_w)$ /J·(mol H$_2$O)$^{-1}$ | $-\Delta_{sol}H_{SO_2}$ /kJ·(mol SO$_2$)$^{-1}$ | $U(\Delta_{sol}H_{SO_2})$ /kJ·(mol SO$_2$)$^{-1}$ |
|---|---|---|---|---|---|---|---|
| | | | | $T$ = 323.15 K | | | |
| 0.36 | 0.0028 | 0.0002 | 0.155 | 64.7 | 23.5 | 23.1 | 8.4 |
| 0.30 | 0.0036 | 0.0004 | 0.200 | 85.3 | 23.5 | 23.7 | 6.5 |
| 0.28 | 0.0043 | 0.0002 | 0.239 | 106.7 | 15.5 | 24.9 | 3.5 |
| 0.29 | 0.0043 | 0.0002 | 0.239 | 130.0 | 15.3 | 30.2 | 3.5 |
| 0.28 | 0.0043 | 0.0006 | 0.239 | 106.7 | 12.5 | 24.8 | 2.9 |
| 0.30 | 0.0047 | 0.0006 | 0.261 | 120.6 | 23.5 | 25.7 | 5.1 |
| 0.32 | 0.0048 | 0.0002 | 0.266 | 137.3 | 16.5 | 28.4 | 3.3 |
| 0.29 | 0.0061 | 0.0003 | 0.339 | 170.2 | 20.0 | 27.7 | 3.3 |
| 0.29 | 0.0064 | 0.0003 | 0.355 | 185.1 | 23.3 | 28.8 | 3.5 |
| 0.28 | 0.0065 | 0.0003 | 0.361 | 168.8 | 21.4 | 26.0 | 3.3 |
| 0.28 | 0.0065 | 0.0008 | 0.361 | 168.8 | 12.7 | 26.0 | 2.0 |
| 0.29 | 0.0086 | 0.0004 | 0.477 | 245.2 | 29.0 | 28.6 | 3.3 |
| 0.28 | 0.0092 | 0.0004 | 0.511 | 232.1 | 26.7 | 25.1 | 2.9 |
| 0.28 | 0.0103 | 0.0004 | 0.572 | 304.8 | 35.1 | 29.5 | 3.3 |
| 0.30 | 0.0112 | 0.0005 | 0.622 | 315.4 | 36.1 | 28.2 | 3.1 |
| 0.29 | 0.0123 | 0.0005 | 0.683 | 335.5 | 39.2 | 27.2 | 3.1 |
| 0.28 | 0.0124 | 0.0005 | 0.688 | 340.2 | 40.0 | 27.5 | 3.3 |
| 0.30 | 0.0135 | 0.0006 | 0.749 | 342.0 | 39.2 | 25.4 | 2.9 |
| 0.30 | 0.0136 | 0.0006 | 0.755 | 391.2 | 48.8 | 28.8 | 3.5 |
| 0.28 | 0.0145 | 0.0006 | 0.805 | 391.3 | 45.1 | 27.0 | 3.1 |
| 0.28 | 0.0166 | 0.0007 | 0.921 | 450.4 | 52.3 | 27.1 | 3.1 |
| 0.30 | 0.0183 | 0.0020 | 1.016 | 496.2 | 25.7 | 27.1 | 1.4 |
| 0.30 | 0.0183 | 0.0008 | 1.016 | 496.2 | 59.8 | 27.1 | 3.3 |
| 0.29 | 0.0185 | 0.0008 | 1.027 | 501.5 | 57.4 | 27.1 | 3.1 |
| 0.29 | 0.0216 | 0.0009 | 1.199 | 552.1 | 63.7 | 25.6 | 2.9 |
| 0.28 | 0.0239 | 0.0010 | 1.327 | 589.2 | 67.4 | 24.7 | 2.7 |
| 0.29 | 0.0259 | 0.0011 | 1.438 | 667.6 | 76.4 | 25.8 | 2.9 |
| 0.29 | 0.0281 | 0.0012 | 1.560 | 693.2 | 79.4 | 24.6 | 2.7 |
| 0.28 | 0.0314 | 0.0013 | 1.743 | 709.4 | 81.1 | 22.6 | 2.5 |
| 0.28 | 0.0335 | 0.0014 | 1.860 | 725.2 | 82.9 | 21.6 | 2.5 |
| 0.28 | 0.0358 [b] | 0.0015 | 1.987 [b] | 735.0 | 84.1 | 20.6 | 2.4 |
| | | | $T = 373.15$ K | | | | |
| 0.28 | 0.0011 | 0.0001 | 0.058 | 21.9 | 1.2 | 20.8 | 1.2 |
| 0.28 | 0.0021 | 0.0003 | 0.117 | 40.6 | 1.2 | 19.3 | 0.6 |
| 0.28 | 0.0024 | 0.0003 | 0.135 | 45.7 | 1.0 | 18.8 | 0.4 |
| 0.28 | 0.0026 | 0.0004 | 0.146 | 48.4 | 1.2 | 18.4 | 0.5 |
| 0.36 | 0.0027 | 0.0003 | 0.152 | 46.7 | 4.4 | 17.1 | 1.6 |
| 0.28 | 0.0032 | 0.0004 | 0.175 | 56.6 | 1.3 | 17.9 | 0.4 |
| 0.28 | 0.0032 | 0.0004 | 0.175 | 56.6 | 1.3 | 17.9 | 0.4 |
| 0.28 | 0.0037 | 0.0005 | 0.204 | 64.6 | 1.3 | 17.5 | 0.3 |
| 0.37 | 0.0042 | 0.0004 | 0.232 | 70.0 | 8.8 | 16.8 | 2.1 |
| 0.28 | 0.0042 | 0.0006 | 0.234 | 71.7 | 1.3 | 17.0 | 0.3 |
| 0.29 | 0.0042 | 0.0006 | 0.234 | 72.1 | 0.9 | 17.1 | 0.2 |
| 0.37 | 0.0050 | 0.0005 | 0.278 | 80.7 | 8.8 | 16.1 | 1.8 |
| 0.37 | 0.0057 | 0.0006 | 0.314 | 88.9 | 8.9 | 15.7 | 1.6 |
| 0.28 | 0.0063 | 0.0008 | 0.350 | 95.1 | 2.5 | 15.1 | 0.4 |
| 0.37 | 0.0065 | 0.0006 | 0.359 | 99.5 | 8.9 | 15.4 | 1.4 |
| 0.36 | 0.0074 | 0.0008 | 0.412 | 103.3 | 8.9 | 13.9 | 1.2 |
| 0.37 | 0.0075 | 0.0008 | 0.419 | 103.6 | 8.9 | 13.7 | 1.2 |
| 0.37 | 0.0084 [b] | 0.0008 | 0.466 [b] | 98.4 | 8.9 | 11.7 | 1.1 |
| 0.37 | 0.0085 [b] | 0.0008 | 0.473 [b] | 95.5 | 7.3 | 11.2 | 0.9 |
| 0.36 | 0.0088 [b] | 0.0009 | 0.488 [b] | 102.5 | 8.9 | 11.7 | 1.0 |
| 0.37 | 0.0098 [b] | 0.0010 | 0.547 [b] | 102.5 | 8.9 | 10.4 | 0.9 |
| 0.37 | 0.0112 [b] | 0.0011 | 0.620 [b] | 108.5 | 8.9 | 9.7 | 0.8 |
| 0.37 | 0.0114 [b] | 0.0011 | 0.634 [b] | 101.7 | 8.1 | 8.9 | 0.7 |
| 0.37 | 0.0126 [b] | 0.0013 | 0.700 [b] | 103.6 | 8.9 | 8.2 | 0.7 |
| 0.37 | 0.0127 [b] | 0.0013 | 0.705 [b] | 107.4 | 8.9 | 8.5 | 0.7 |
| 0.37 | 0.0142 [b] | 0.0014 | 0.786 [b] | 101.3 | 8.9 | 7.2 | 0.6 |

[a] Standard uncertainties: $u(T) = 0.03$ K; $u(p) = 0.02$ MPa; $u(\bar{m}_{SO_2}) = u(\alpha)/M_w$ (where $M_w$ is the molar mass of water).

[b] Value of $\alpha$ above the gas solubility limit calculated by the model. The point is in a two-phase region; $\alpha$ and $\bar{m}_{SO_2}$ correspond to the total composition of the two-phase system.

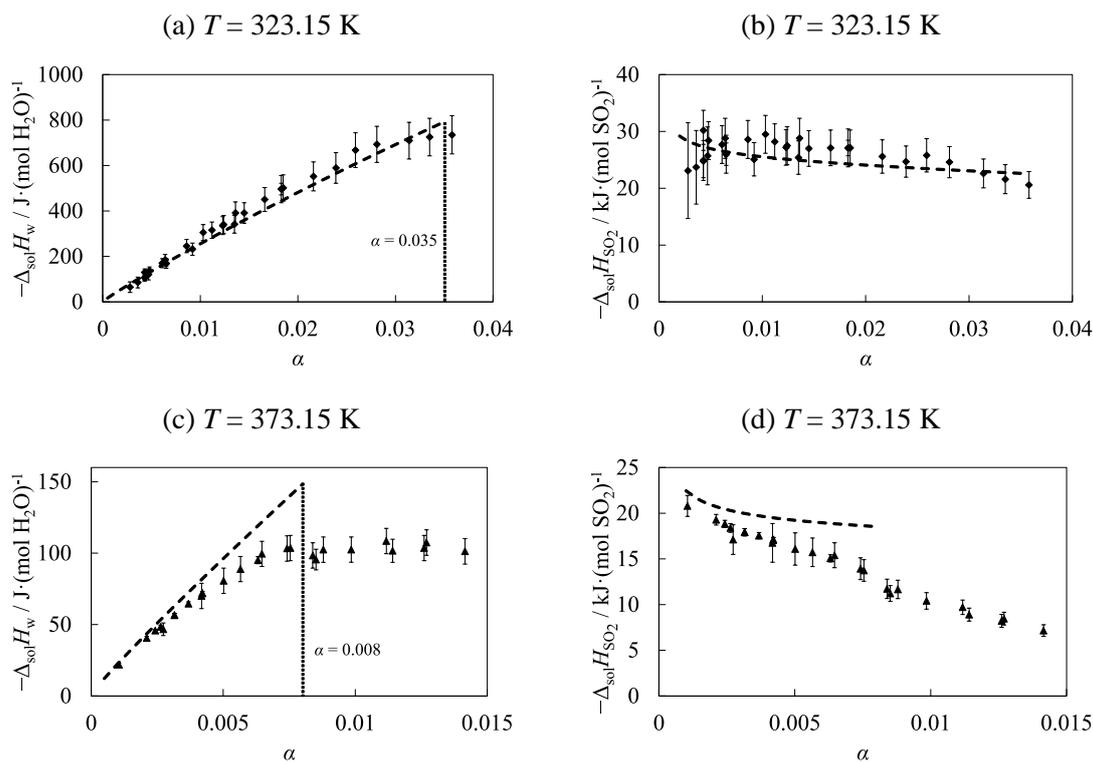

Figure 4. Enthalpy of solution of gaseous sulfur dioxide in liquid water per mole of water (a, c), $\Delta_{sol}H_w$, and per mole of sulfur dioxide (b, d), $\Delta_{sol}H_{SO_2}$, at average pressure $<p> = 0.31$ MPa and temperature $T$. (♦) and (▲): experimental values at 323.15 K and 373.15 K (this work); (- - -): calculated values. Vertical dotted line: solubility limit calculated by the model. Error bars: expanded uncertainty (95% confidence level).

The speciation is calculated by the model as function of gas loading charge. It shows that only $SO_2$ and $HSO_3^-$ are significantly present in solution (Figure 5); the molalities of $SO_3^{2-}$ and $OH^-$ species at equilibrium are comparable and negligible. The enthalpy of solution $\Delta_{sol}H_{SO_2}$ is calculated as a combination of four contributions: $\Delta_{r,I}H$ for $HSO_3^-$ formation, $\Delta_{r,II}H$ for $SO_3^{2-}$ formation, $\Delta_{r,w}H$ for water dissociation and $\Delta_{r,8}H$ for physical dissolution. In agreement with the speciation calculations, the only non-negligible enthalpy contributions are issued from $HSO_3^-$ formation and physical dissolution (Figure 6), the latter being the dominant contribution. Basically, the mechanism of $SO_2$ dissolution in water is depicted by the model as a physical dissolution in water followed by a chemical dissociation yielding $HSO_3^-$ formation. At 373.15 K, the overestimated enthalpy of solution could indicate that the chemical dissociation is less significant than the one estimated by the model. Indeed, chemical dissolution being more energetic than physical dissolution, this mechanism will increase the calculated total enthalpy of solution.

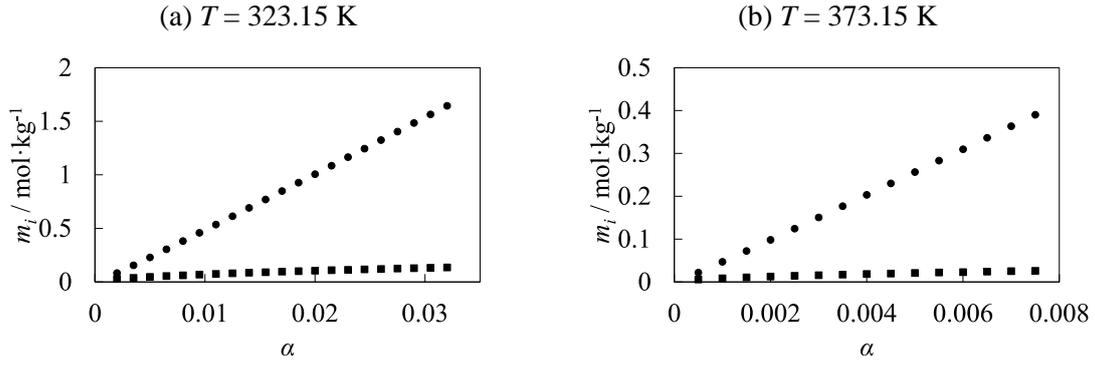

Figure 5. Equilibrium molality, $m_i$, of species $i$ present in the liquid phase after the dissolution of gaseous sulfur dioxide in liquid water at temperature $T$ as functions of the loading charge, $\alpha$. (●): $SO_2$; (■): $HSO_3^-$ and $H^+$.

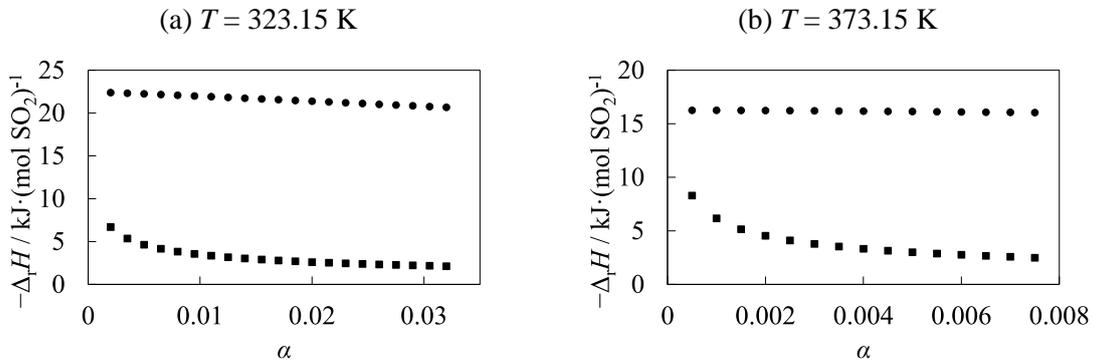

Figure 6: Model contributions to the enthalpy of solution of gaseous sulfur dioxide in liquid water per mol of $SO_2$, $\Delta_{sol}H_{SO_2}$, at temperature $T$ as functions of the loading charge, $\alpha$. (●): physical dissolution ($\Delta_{r,8}H$); (■): $HSO_3^-$ formation ($\Delta_{r,I}H$).

## 4. {NO-water} system

### 4.1. Thermodynamic representation

Considering the low solubility of NO in water, the liquid solution was assumed to be ideal and the vapor-liquid equilibrium was represented by Henry's law. Experimental data on the dissolution of NO in aqueous solution are scarce. The solubility of NO in water has been investigated by Winkler [22] at a partial pressure of NO ($p_{NO}$) of 101325 Pa and temperatures from 273 to 353 K. Armor [23] has studied the influence of the pH and the ionic strength on the solubility of NO in aqueous solutions. These data were used by Shaw and Vosper [36] to calculate Henry's law constants. Several reviews have been published later [37-40] based on the experimental data from Winkler [22] and Armor [23]. These reviews propose different equations for the solubility limit and the Henry's constant as functions of temperature. The mole-fraction based Henry's law constants at $p_{NO}$ = 101325 Pa ($H_{NO,w}$) were here obtained from the equation reported by Gevantman [39] for solubility as a function of temperature and the same partial pressure of NO, and are represented by eq. 28, where $A$ = 179.418, $B$ = –8234.20 K, and $C$ = –22.8155.

$$\ln\left(H_{NO,w}/Pa\right) = A + \frac{B}{T} + C \cdot \ln(T/K) \tag{28}$$

The enthalpy of solution was estimated from Henry's law constant, neglecting its pressure dependence and using Van't Hoff equation (eq. 29).

$$\Delta_{sol}H = R(B - C \cdot T) \tag{29}$$

### 4.2. Results

The calorimetric measurements for NO dissolution in water were carried out at temperature $T$ = 323.15 K and pressures around the average values 2.27 MPa, 2.56 MPa and 2.80 MPa. The enthalpies of solution of NO in water were measured at 323.15 K. Due to the low solubility of NO in water at 373.15 K, it was not possible to perform reliable enthalpy of solution experiments at this temperature. The experimental values are reported in Table 7. The measurements were carried out with a loading charge above the gas solubility limit estimated from literature data and, accordingly, they have been considered to correspond to a two-phase system. The loading charge and overall molality refer to the total composition of the two-phase system. The experimental and calculated enthalpies of solution are illustrated in Figure 7.

Table 7. Enthalpy of solution of gaseous nitrogen monoxide in liquid water per mole of water, $\Delta_{sol}H_w$, and per mole of nitrogen monoxide, $\Delta_{sol}H_{NO}$, at temperature $T$ = 323.15 K and pressure $p$ (final pressure after mixing), as functions of the loading charge of the gas (moles of gas per 1 mole of water), $\alpha = M_w \bar{m}_{NO}$ ($M_w$, molar mass of water; $\bar{m}_{NO}$, overall molality of NO). $u(X)$, standard uncertainty of the quantity $X$; $U(X)$ expanded uncertainty (95% confidence level) of $X$. Measurements are carried out around an average pressure $<p>$. [a]

| $p$ /MPa | $\alpha$ [b] | $u(\alpha)$ | $\bar{m}_{NO}$ [b] /mol·kg$^{-1}$ | $-\Delta_{sol}H_w$ /J·(mol H$_2$O)$^{-1}$ | $U(\Delta_{sol}H_w)$ /J·(mol H$_2$O)$^{-1}$ | $-\Delta_{sol}H_{NO}$ /kJ·(mol NO)$^{-1}$ | $U(\Delta_{sol}H_{NO})$ /kJ·(mol NO)$^{-1}$ |
|---|---|---|---|---|---|---|---|
| | | | | $<p>$ = 2.80 MPa | | | |
| 2.80 | 0.00201 | 0.00009 | 0.112 | 4.1 | 0.6 | 2.03 | 0.25 |
| 2.80 | 0.00403 | 0.00017 | 0.224 | 5.6 | 1.0 | 1.38 | 0.25 |
| 2.80 | 0.00403 | 0.00017 | 0.224 | 5.3 | 0.8 | 1.32 | 0.18 |
| 2.80 | 0.00604 | 0.00026 | 0.335 | 6.3 | 1.0 | 1.05 | 0.16 |
| | | | ⟨p⟩ = 2.56 MPa | | | | |
| 2.55 | 0.00092 | 0.00004 | 0.051 | 2.5 | 0.6 | 2.74 | 0.71 |
| 2.56 | 0.00184 | 0.00008 | 0.102 | 4.6 | 0.6 | 2.49 | 0.31 |
| 2.56 | 0.00276 | 0.00012 | 0.153 | 5.1 | 0.6 | 1.85 | 0.22 |
| 2.56 | 0.00368 | 0.00016 | 0.204 | 6.1 | 0.8 | 1.66 | 0.20 |
| 2.56 | 0.00461 | 0.00020 | 0.256 | 6.1 | 0.8 | 1.32 | 0.16 |
| 2.56 | 0.00552 | 0.00023 | 0.306 | 6.3 | 0.8 | 1.15 | 0.14 |
| 2.56 | 0.00645 | 0.00027 | 0.358 | 6.1 | 1.0 | 0.95 | 0.14 |
| 2.57 | 0.00737 | 0.00031 | 0.409 | 6.1 | 0.8 | 0.83 | 0.10 |
| | | | ⟨p⟩ = 2.27 MPa | | | | |
| 2.27 | 0.00081 | 0.00003 | 0.045 | 4.6 | 0.6 | 5.62 | 0.69 |
| 2.28 | 0.00082 | 0.00003 | 0.046 | 4.2 | 0.6 | 5.11 | 0.76 |
| 2.27 | 0.00163 | 0.00007 | 0.090 | 5.5 | 0.6 | 3.35 | 0.39 |
| 2.27 | 0.00164 | 0.00007 | 0.091 | 5.3 | 0.6 | 3.22 | 0.39 |
| 2.28 | 0.00245 | 0.00010 | 0.136 | 6.1 | 0.8 | 2.47 | 0.29 |
| 2.27 | 0.00245 | 0.00010 | 0.136 | 5.5 | 0.6 | 2.23 | 0.27 |
| 2.27 | 0.00326 | 0.00014 | 0.181 | 5.5 | 0.8 | 1.67 | 0.24 |
| 2.28 | 0.00328 | 0.00014 | 0.182 | 5.8 | 0.8 | 1.77 | 0.22 |
| 2.27 | 0.00408 | 0.00017 | 0.226 | 5.6 | 0.6 | 1.36 | 0.16 |
| 2.28 | 0.00410 | 0.00017 | 0.228 | 6.0 | 0.8 | 1.47 | 0.18 |
| 2.28 | 0.00491 | 0.00021 | 0.273 | 6.2 | 0.8 | 1.26 | 0.16 |
| 2.29 | 0.00491 | 0.00021 | 0.273 | 6.3 | 0.8 | 1.29 | 0.16 |
| 2.28 | 0.00573 | 0.00024 | 0.318 | 6.6 | 0.8 | 1.16 | 0.14 |
| 2.28 | 0.00654 | 0.00028 | 0.363 | 6.1 | 1.0 | 0.94 | 0.14 |
| 2.27 | 0.00831 | 0.00035 | 0.461 | 6.1 | 0.8 | 0.74 | 0.10 |

[a] Standard uncertainties: $u(T) = 0.03$ K; $u(p) = 0.02$ MPa; $u(\bar{m}_{NO}) = u(\alpha)/M_w$ (where $M_w$ is the molar mass of water).

[b] Values of $\alpha$ are above the gas solubility limit estimated from literature data. The points are considered to be in a two-phase region; $\alpha$ and $\bar{m}_{NO}$ correspond to the total composition of the two-phase system.

The plot in Figure 7a shows that the enthalpy of solution $\Delta_{sol}H_w$ increases with the gas loading charge $\alpha$ up to the solution saturation and remains constant within experimental uncertainty when no more gas can be absorbed. The intersection between the two parts of the curve should correspond to the limit of gas solubility.

A theoretical gas solubility limit ($\alpha_{calc} = 6.45 \cdot 10^{-4}$) at 323.15 K and 2.5 MPa was estimated using the value at atmospheric pressure ($\alpha = 2.58 \cdot 10^{-5}$) reported by Gevantman [39] and Henry's law constant (eq. 28). This calculated solubility is represented in Figure 7a, together with the plot of a hypothetical enthalpy behavior, i.e. a linear increase up to reach a plateau. This limit of solubility is lower than those which can be estimated experimentally. For gas loading charge below 0.00245, the measured enthalpy of solution is lower than expected. However, increasing slightly the pressure from 2.27 to 2.80 MPa the enthalpies get closer to the hypothetical plateau. The gap between experimental values and the plateau could be due to mixing problems when running experiments at very low gas flow rates (i.e. below 0.05 mL/min).

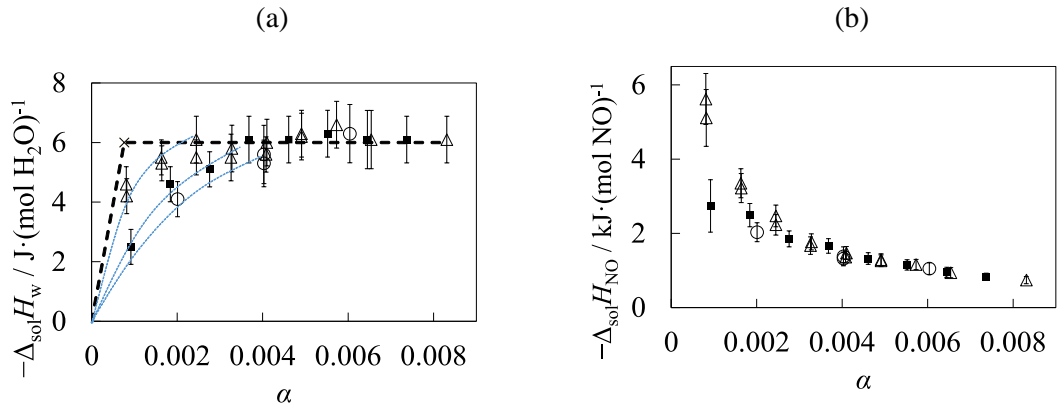

Figure 7. Enthalpy of solution of gaseous nitrogen monoxide in liquid water per mole of water (a), $\Delta_{sol}H_w$, and per mole of nitrogen monoxide (b), $\Delta_{sol}H_{NO}$, at temperature $T = 323.15$ K and average pressure $<p>$. (△): $<p>$ = 2.27 MPa; (■): $<p>$ = 2.56 MPa; (○): $<p>$ = 2.80 MPa; (✗): estimated solubility limit; (- - -) : hypothetical enthalpy behavior; dotted lines: curves connecting the points at constant $<p>$ = 2.27 MPa, 2.50 MPa or 2.80 MPa. Error bars: expanded uncertainty (95% confidence level).

Because of the lack of reliable experimental data points in the unsaturated domain (Figure 7), the enthalpy of solution per mole of gas, $\Delta_{sol}H_{NO}$ can only be estimated using the average enthalpy value on the plateau ($\Delta_{sol}H_w$ = (–6.0 ± 0.5) J·(mol $H_2O$)$^{-1}$) and the theoretically estimated limit of gas solubility ($\alpha_{calc} = 6.45 \cdot 10^{-4}$). Assuming a linear increase of the enthalpy up to the plateau, $\Delta_{sol}H_{NO}$ is estimated to be (–9.5 ± 0.8) kJ·(mol NO)$^{-1}$; the calculated $\Delta_{sol}H_{NO}$ value at 323.15 K and atmospheric pressure using literature Henry's constants (eq. 28-29) is (–7.2 ± 0.2) kJ·(mol NO)$^{-1}$. More experimental measurements at low gas loading charges will be necessary to conclude on the consistency between solubility data and enthalpy data.

## 5. Conclusion

Dissolution of $SO_2$ and NO in water was investigated using a thermodynamic approach. Experimental enthalpies of solution were determined as a function of temperature and pressure, and their consistency with literature solubility data was tested. The dissolution of $SO_2$ in water was described using a thermodynamic model representative of vapor-liquid equilibrium and gas chemical dissociations. The enthalpies of solution were derived from the model as a combination of terms related to physical dissolution and gas dissociation. The main contribution to the enthalpy of solution has been found to be the physical dissolution. At 373.15 K, the model predicts accurately the solubility limits but slightly overestimates the enthalpy of solution. For dissolution of NO in water, the system exhibits very low gas solubility. It was difficult to measure enthalpies of solution for a gas loading charge below the limit of solubility. Combining our result with literature values of solubility and Henry's constants, the experimental enthalpy data at 323.15 K seem consistent. However, the calorimetric technique used in this work will require some improvement to investigate the domain where the solution is unsaturated and make it possible to develop a rigorous model representative of vapor-liquid equilibrium and enthalpy of solution.

## Acknowledgements

F. Hevia is grateful to Ministerio de Educación, Cultura y Deporte for the grant FPU14/04104 and for the complementary grants EST16/00824 and EST17/00292.

This work was realized with the financial support of French National Agency for Research, through the program SIGARRR (ANR-13-SEED-0006).